\documentstyle{mn}
 
\newif\ifAMStwofonts 
 
\def\today{\ifcase\month\or
 January\or February\or March\or April\or May\or June\or
 July\or August\or September\or October\or November\or
 December\fi\space\number\day, \number\year}

\def\todmy{\number\day\space\ifcase\month\or
 January\or February\or March\or April\or May\or June\or
 July\or August\or September\or October\or November\or
 December\fi\space\number\year}

\newcommand{\bdisp} {\begin{displaymath}}
\newcommand{\edisp} {\end{displaymath}}
\newcommand{\beqn} {\begin{equation}}
\newcommand{\eeqn} {\end{equation}}
\newcommand{\beqr} {\begin{array}}
\newcommand{\eeqr} {\end{array}}

\newcommand{\tal}{\it et al. \rm}

\ifoldfss 
  \ifCUPmtlplainloaded \else 
    \NewTextAlphabet{textbfit} {cmbxti10} {} 
    \NewTextAlphabet{textbfss} {cmssbx10} {} 
    \NewMathAlphabet{mathbfit} {cmbxti10} {} 
    \NewMathAlphabet{mathbfss} {cmssbx10} {} 
  \fi 
  \ifAMStwofonts 
    \ifCUPmtlplainloaded \else 
      \NewSymbolFont{upmath} {eurm10} 
      \NewSymbolFont{AMSa} {msam10} 
      \NewMathSymbol{\upi}     {0}{upmath}{19} 
      \NewMathSymbol{\umu}     {0}{upmath}{16} 
      \NewMathSymbol{\upartial}{0}{upmath}{40} 
      \NewMathSymbol{\leqslant}{3}{AMSa}{36} 
      \NewMathSymbol{\geqslant}{3}{AMSa}{3E}

    \fi 
  \fi 
\fi 
 
\ifnfssone 
  \newmathalphabet{\mathit} 
  \addtoversion{normal}{\mathit}{cmr}{m}{it} 
  \addtoversion{bold}{\mathit}{cmr}{bx}{it} 
  \newmathalphabet{\mathbfit} 
  \addtoversion{normal}{\mathbfit}{cmr}{bx}{it} 
  \addtoversion{bold}{\mathbfit}{cmr}{bx}{it} 
  \newmathalphabet{\mathbfss} 
  \addtoversion{normal}{\mathbfss}{cmss}{bx}{n} 
  \addtoversion{bold}{\mathbfss}{cmss}{bx}{n} 
  \ifAMStwofonts 
    \ifCUPmtlplainloaded \else 
      \UseAMStwoboldmath 
      \makeatletter 
      \new@mathgroup\upmath@group 
      \define@mathgroup\mv@normal\upmath@group{eur}{m}{n} 
      \define@mathgroup\mv@bold\upmath@group{eur}{b}{n} 
      \edef\UPM{\hexnumber\upmath@group} 
      \new@mathgroup\amsa@group 
      \define@mathgroup\mv@normal\amsa@group{msa}{m}{n} 
      \define@mathgroup\mv@bold\amsa@group{msa}{m}{n} 
      \edef\AMSa{\hexnumber\amsa@group} 
      \makeatother 
      \mathchardef\upi="0\UPM19 
      \mathchardef\umu="0\UPM16 
      \mathchardef\upartial="0\UPM40 
      \mathchardef\leqslant="3\AMSa36 
      \mathchardef\geqslant="3\AMSa3E 
    \fi 
  \fi 
\fi 
 
\ifnfsstwo 
  \DeclareMathAlphabet{\mathbfit}{OT1}{cmr}{bx}{it} 
  \SetMathAlphabet\mathbfit{bold}{OT1}{cmr}{bx}{it} 
  \DeclareMathAlphabet{\mathbfss}{OT1}{cmss}{bx}{n} 
  \SetMathAlphabet\mathbfss{bold}{OT1}{cmss}{bx}{n} 
  \ifAMStwofonts 
    \ifCUPmtlplainloaded \else 
      \DeclareSymbolFont{UPM}{U}{eur}{m}{n} 
      \SetSymbolFont{UPM}{bold}{U}{eur}{b}{n} 
      \DeclareSymbolFont{AMSa}{U}{msa}{m}{n} 
      \DeclareMathSymbol{\upi}{0}{UPM}{"19} 
      \DeclareMathSymbol{\umu}{0}{UPM}{"16} 
      \DeclareMathSymbol{\upartial}{0}{UPM}{"40} 
      \DeclareMathSymbol{\leqslant}{3}{AMSa}{"36} 
      \DeclareMathSymbol{\geqslant}{3}{AMSa}{"3E} 
    \fi 
  \fi 
\fi 
 
\ifCUPmtlplainloaded \else 
  \ifAMStwofonts \else 
    \def\upi{\pi} 
    \def\umu{\mu} 
    \def\upartial{\partial} 
  \fi 
\fi 

\begin{document}

\title{Evolution of Compact Groups of Galaxies I. Merging rates}
\author[E. Athanassoula, J. Makino, A. Bosma]
       {E. Athanassoula,$^1$, J. Makino,$^2$ and A. Bosma$^1$\\ 
       $^1$Observatoire de Marseille,
       2 Place Le Verrier,
       F-13248 Marseille Cedex 4, France\\
       $^2$Department of Information Science and Graphics,
College of Arts and Sciences, The University of Tokyo,
Tokyo 153, Japan}

\date{Accepted .
      Received ;
      in original form 1996 June}

\pagerange{\pageref{firstpage}--\pageref{lastpage}}
\pubyear{1996}
\maketitle

\label{firstpage}

\begin{abstract}
We discuss the merging rates in compact groups of 5 identical elliptical
galaxies. All groups have the same mass and binding energy. We consider both 
cases with individual halos and cases where the halo is common
to all galaxies and enveloping the whole group. In the latter 
situation the merging rate is slower if 
the halo is more massive. The mass of individual halos has little influence 
on the merging rates, due to the fact that all galaxies in our simulations
have the same mass, and so the more extended ones have a smaller velocity 
dispersion. Groups with individual halos merge faster than groups with common 
halos if the configuration is centrally concentrated, like a King distribution
of index $\Psi=10$. On the other hand for less concentrated configurations the
merging is initially faster for individual halo cases, and slower after part 
of the group has merged. In cases with common halo, centrally concentrated 
configurations merge faster for high halo-to-total mass ratios and slower for 
low halo-to-total mass ratios. Groups whose virial ratio is initially less 
than one merge faster, while groups that have initially cylindrical rotation 
merge slower than groups starting in virial equilibrium.  

In order to test how long a virialised group can survive before merging we 
followed the evolution of a group with a high halo-to-total mass ratio and a 
density distribution with very little central concentration. We find that the 
first merging occurred only after a large number of crossing times. A 
reasonable calibration of our computer units shows that this time should be 
larger than a Hubble time. Therefore, our simulations suggest that, at least 
for appropriate initial conditions, the longevity of compact groups is not 
necessarily a problem, thus presenting an alternative explanation to why we 
observe so many compact groups despite the fact that their lifetimes seem 
short. 
\end{abstract}

\section{Introduction}
\indent
Hickson compact groups (hereafter HCGs) are tight associations of at least 
four ga\-la\-xies fulfilling specific criteria concerning their compactness and
isolation. Hickson (1982), using the Palomar
sky survey prints, catalogued 100 such systems of which 92 were subsequently
found to contain at least three members with accordant redshifts (Hickson 
\tal 1992). The observed projected separations imply very high space densities,
equal to or greater than those found in the cores of clusters
of galaxies, while their
velocity dispersions are moderate (Hickson \tal 1992), of the same order as 
velocity dispersions 
in bright galaxies. This makes them ideal sites for interactions and 
mergings. Thus the question of why such groups are still observed and have
not merged already arises naturally.
Either for some reason the relevant time-scales for their merging 
are longer than expected, or compact groups are a relatively short lived
evolutionary phase in a frequently occurring scenario, or they are not real 
configurations but chance projections. 

Barnes (1985) considers the possibility that a common dark halo might envelop a
compact group and shows with the help of N body simulations that this slows
down the merging process. His groups are constituted of 5 to 10 elliptical
galaxies and each galaxy is represented by 75 or 100 mass points. His 
conclusion has
been corroborated by the work of Navarro, Mosconi \& Garcia 
Lambas (1987) and Cavaliere \tal (1982),
although those simulations do not strictly pertain to compact groups and are
made with much fewer particles. More recently Bode, Cohn \& Lugger (1992), 
using  a total
of 5000 particles and King laws for all mass distributions (mass within a given
galaxy, distribution of galaxies in the group and distribution of mass within
the common halo), come to the same conclusion as Barnes. On the other
hand Athanassoula and  Makino (1995), using Plummer laws for all
the mass distributions, found that the complete merging of the group into 
one object happens earlier in cases with common halos than in cases of 
individual halos with the same distribution as the light component.

Thus, although the amount and distribution of dark will 
influence the merging times, it is not clear exactly how. Other factors such 
as the distribution of the galaxies in the group, or their kinematics, have 
not 
yet been fully investigated. In this paper we  
return to the problem, using a larger number of particles
and, in particular, trying more simulations with different distributions, in
order to understand if and how these distributions influence
the merging rates. Our simulations are presented in section \ref{sec:simul}
and the criterion for merging in section \ref{sec:rates}. Some general 
information on the evolution of the runs is given in section \ref{sec:results}.
Merging rates are compared in section \ref{sec:crates}. Section 
\ref{sec:pairs} discusses pairs and triplets.  A general discussion is given 
in section 
\ref{sec:discuss}, where we also present one simulation which takes a long 
time before merging. 

\section{Simulations}
\label{sec:simul}
\indent

\subsection{Galaxy models}
\label{sec:galmod}
\indent
To simplify the problem, we have modelled all galaxies constituting a group
as ellipticals. This results in savings on both the number of particles and the
amount of simulations to be performed, since more particles are necessary to
model disc galaxies, and the relative orientation of these discs can
influence the merging rates (White 1979, Barnes 1992).  Furthermore all 
galaxies
constituting the group are taken to be identical (simulations of groups with
unequal mass galaxies have been reported by Governato, Bhatia \& Chincarini 
(1991), by Weil \& Hernquist 1994 and 1996, and by
Athanassoula \& Makino (1995)). Thus our
individual galaxy models will represent ellipticals with or without individual
halos, and different masses and sizes will be considered for the halo
component. To enable meaningful comparisons, we have ensured that the
radial dependences of the density of the luminous material of all the galaxy 
models are as similar as possible.
This is of course only approximate, since a figure in equilibrium in a halo
is not the same as a figure in equilibrium in its own gravitational
potential. Our setup procedure, however, ensures considerable similarity. 

The galaxies are modelled as Plummer spheres or, in the case of galaxies with
halos, as composite models represented by two superposed Plummer spheres which
are evolved together to equilibrium\footnote{An alternative, and certainly more
elegant, way of constructing composite models is to calculate, if possible, 
analytical distribution functions of such systems, as done recently by Ciotti 
(1996) for a Hernquist model in a Hernquist halo}. More specifically, we have 
used five 
galaxy models, denoted in Tables 1, 2 and 3 by g0, g1, g2, g3 and g4. 
Model g0 represents a halo-less elliptical and is mo\-del\-led by a
simple Plummer sphere. Model g1 represents an elliptical
with a halo of relatively short extent and has been 
constructed in the following way:
We create two Plummer spheres of equal mass, each with 2048 particles. We 
rescale one in radius by a factor of 0.5 and call, for
simplicity, the extended component the halo and the less extended one the 
luminous part.
Then we evolve the composite configuration for a time interval 
sufficient to reach equilibrium. Keeping its energy constant, we now 
rescale this model so that the 
half-mass radius of the luminous part is equal to that of the simple 
Plummer model and its mass equal to that of the simple Plummer model g0 
used in 
simulations with common halos and the same halo-to-total mass ratio.

Model g2 
was built in exactly the same way as g1 except that the ratio of the extents 
of the two components is taken to be 0.25, instead of 0.5, thus aiming for a 
halo of
larger extent compared to the optical part. Model g3 has a halo which is three
times more massive than the luminous part and is initially four times as
extended. Finally the halo of model g4 is seven times more massive than 
the optical part and is initially eight times more extended. Before being 
used in the group simulations, all composite models were 
evolved for 30 time units, an ample time for equilibrium between the 
two components to be reached. 

\begin{table}
\label{tab:indgal1}
\caption{Global parameters for individual galaxies}
\begin{center}
\begin{tabular}{|c|c|c|c|c|c|c|c|c|c|c|c|}
\hline
galaxy & $M_{ch}/M_t$ & $M_{ih}/M_t$ & $E$ & $\sqrt{2|E|/M}$ \\
\hline
\hline
g0 & 0.5 & 0.& -0.98 & 0.99  \\
g0 & 0.75 & 0. & -0.24 & 0.69  \\
g0 & 0.875 & 0. & -0.06 & 0.50  \\
g1 & 0. & 0.5 & -2.71 & 1.16  \\
g2 & 0. & 0.5 & -2.20 & 1.05 \\
g3 & 0. & 0.75 & -1.49 & 0.86 \\
g4 & 0. & 0.875 & -0.82 & 0.64 \\
\hline
\end{tabular}
\end{center}
\end{table}
\begin{table}
\label{tab:indgal2}
\caption{Radii for individual galaxies}
\begin{center}
\begin{tabular}{|c|c|c|c|c|c|c|c|c|c|c|c|}
\hline
galaxy & $r(.3)$& $r(.8)$ & $r_{l}(.3)$ 
& $r_{l}(.8)$ & $r_{h}(.3)$ & $r_{h}(.8)$ \\
\hline
\hline
g0 & 0.5 & 1.5 & 0.5 & 1.5 & --- & --- \\
g1 & 0.7 & 2.7 & 0.5 & 1.7 & 1.0 & 3.6 \\
g2 & 0.8 & 3.9 & 0.5 & 1.6 & 1.6 & 6.1 \\
g3 & 1.2 & 5.5 & 0.5 & 1.5 & 1.9 & 6.5 \\
g4 & 2.4 & 9.4 & 0.5 & 1.6 & 3.1 & 10.\\
\hline
\end{tabular}
\end{center}
\end{table}

Table 1 gives some properties of the galaxies we have
used. Column 1 gives the name of the galaxy, in the notation introduced 
above, columns 2 and 3 its halo-to-total mass ratio (for the common and 
individual halos respectively), column 4 its binding
energy and column 5 a measure of its velocity dispersion. The 
density profiles of the evolved ga\-la\-xies are compared in Table 
2. Columns 1 and 2 give 
the radii containing 30\% and 80\% of the total galaxy mass, columns 3 and 4 
the same radii for the luminous mass, and columns 5 and 6
for the halo. We note that the luminous parts of the three configurations are 
quite
similar, as desired. The total density distribution of the models
with halos is consi\-de\-rably more extended than that of model g0, while
models g1 and g2 differ mainly in their outer parts. 
We note that the characteristic radii of composite models are somewhat 
smaller than those of their unevolved halos, as a result of the 
additional pull 
inwards from the luminous part.

\subsection{Models for the distribution of galaxies in the group and for
the common halo}
\label{sec:grouphal}
\indent
Each group is initially constituted of 5 identical galaxies whose initial
positions and velocities are drawn randomly
from a distribution which is either a
Plummer model or a King model of a given index. Only King models with indexes
$\Psi=1, 5$ or 10 have been used. The 
concentration of the King models increases with their index $\Psi$. Thus the 
$\Psi=10$ distribution is the most 
centrally concentrated one, while having an extended outer part.
The intermediate case $\Psi=5$ 
is rather similar to the Plummer distribution. 
Five realisations of each of these four
distributions have been made and used identically both for the simulations with
common halos and for the ones without. 

The models for the common
halo distribution have  also been chosen to be either
Plummer models or King models with indexes 1, 5 or 10. For simplicity, and to 
keep the number of simulations limited, in all our models the halo
mass is either in individual galactic halos or in common halos. No cases with
part of the mass in one type of halo and part of the mass in the other have
been considered. Three different fractions of halo mass have been considered,
corresponding to $M_{lum}/M_{tot}=$ 0.5, 0.25 or 0.125, where $M_{lum}$ is the
mass in the luminous parts of the galaxies and $M_{tot}$ is the total mass in
the simulation. 

	Most of our simulations start off with a virial ratio $2T/|W|$ equal 
to 1 and random orientations of the ve\-lo\-ci\-ty vectors. Nevertheless 
the models where the distribution for the
group and the halo are not of the same type will start off equilibrium. We
have also run 70 simulations of Plummer groups in (whenever relevant)
Plummer halos with different initial kinematics. For reasons that
will be discussed in the following sections, all but five had a
$M_{lum}/M_{tot}=0.125$, the remaining five having a $M_{lum}/M_{tot}=0.5$.
We experimented with cylindrically rotating, cold and expanding groups. For
the rotating ones we proceed as follows. We first create the group in
virial equilibrium and then add rotation in the $(x,y)$ plane by setting the
velocity in that plane to be perpendicular to the position vector in that
plane. This means that we give these models the maximum possible velocity
around the $z$ axis without any change in their $z$-component of velocity and
their kinetic energy. Cold groups have a virial ratio $2T/|W|=0.25$ or 
0.5, while 
having the same total energy and mass as the ones in virial equilibrium. 
Finally for the expanding groups the galaxy bulk-velocity vectors are 
oriented such 
that there is only expansion, their amplitude is taken to be proportional to 
the distance of the galaxy from the center of the group,  while the total mass
and energy are kept constant. The same 
procedure is also followed, whenever relevant, for the common halo, except 
that now we eliminate the halo particles that are unbound.

\begin{table*}
\centering
\begin{minipage}{140mm}
\label{tab:listruns}
\caption{List of runs}
\begin{center}
\begin{tabular}{|c|c|c|c|c|c|c|c|}
\hline
Run  & Galaxies & Group & $CH$ & $M_{halo}/M_{tot}$ & & $N_g$ & $N_s$ \\
\hline
\hline
pl54, pl47, pl48, pl49, pl50 & g1 & p & - & 0.5 & $virial$ & 4096 & 20480 \\
pl55, pl17, pl51, pl52, pl53 & g2 & p & - &  0.5 & $virial$ & 4096 & 20480 \\
pl56, pl22, pl23, pl24, pl25 & g0 & p & p & 0.75 & $virial$ & 1635 & 32700 \\
pl27, pl28, pl29, pl30, pl31 & g3 & p & - & 0.75 & $virial$ & 4096 & 20480 \\
pl37, pl38, pl39, pl40, pl41 & g0 & p & p & 0.5 & $virial$ & 1635 & 16350 \\
pl42, pl43, pl44, pl45, pl46 & g0 & p & p & 0.875 & $virial$ & 1635 & 65400 \\
pl57, pl58, pl59, pl60, pl61 & g4 & p & - & 0.875 & $virial$ & 4096 & 20480 \\
ki2, ki3, ki4, ki5, ki6 & g2 & k5 & - & 0.5 & $virial$ & 4096 & 20480 \\
ki7, ki8, ki9, ki10, ki11 & g1 & k5 & - & 0.5 & $virial$ & 4096 & 20480 \\
ki12, ki13, ki14, ki15, ki16 & g0 & k5 & k5 & 0.75 & $virial$ & 1635 & 32700 \\
ki17, ki18, ki19, ki20, ki21 & g0 & k10 & k10 & 0.75 & $virial$ & 1635 & 32700 \\
ki22, ki23, ki24, ki25, ki26 & g2 & k10 & - & 0.5 & $virial$ & 4096 & 20480 \\
ki27, ki28, ki29, ki30, ki31 & g1 & k10 & - & 0.5 & $virial$ & 4096 & 20480 \\
ki32, ki33, ki34, ki35, ki36 & g0 & k10 & k10 & 0.5 & $virial$ & 1635 & 16350 \\
ki37, ki38, ki39, ki40, ki41 & g3 & k10 & - & 0.75 & $virial$ & 4096 & 20480 \\
ki47, ki48, ki49, ki50, ki51 & g0 & k10 & k10 & 0.875 & $virial$ & 1635 & 65400 \\
ki52, ki53, ki54, ki55, ki56 & g4 & k10 & - & 0.875 & $virial$ & 4096 & 20480 \\
ki59, ki60, ki61, ki62, ki63 & g4 & k1 & - & 0.875 & $virial$ & 4096 & 20480 \\
ki64, ki65, ki66, ki67, ki68 & g0 & k1 & k1 & 0.875 & $virial$ & 1635 & 65400 \\
ki69, ki70, ki71, ki72, ki73 & g0 & k1 & k1 & 0.5 & $virial$ & 1635 & 16350 \\
ki74, ki75, ki76, ki77, ki78 & g3 & k1 & - & 0.75 & $virial$ & 4096 & 20480 \\
ki79, ki80, ki81, ki82, ki83 & g0 & k1 & k1 & 0.75 & $virial$ & 1635 & 32700 \\
ki84, ki85, ki86, ki87, ki88 & g1 & k1 & - & 0.5 & $virial$ & 4096 & 20480 \\
kp2, kp3, kp4, kp5, kp6 & g0 & k10 & p & 0.75 & $virial$ & 1635 & 32700 \\
kp7, kp8, kp9, kp10, kp11 & g0 & k10 & p & 0.5 & $virial$ & 1635 & 16350 \\
pk5, pk6, pk12, pk13, pk14 & g0 & p & k10 & 0.75 & $virial$ & 1635 & 32700 \\
pk10, pk11, pk15, pk16, pk17 & g0 & p & k10 & 0.5 & $virial$ & 1635 & 16350 \\
\hline
pl62, pl63, pl64, pl65, pl66 & g4 & p & - & 0.875 & $rotating$ & 4096 & 20480 \\
pl67, pl68, pl69, pl70, pl71 & g0 & p & p & 0.875 & $rotating$ & 1635 & 65400 \\
pl77, pl78, pl79, pl80, pl81 & g0 & p & p & 0.875 & $cold$ & 1635 & 65400 \\
pl82, pl83, pl84, pl85, pl86 & g4 & p & - & 0.875 & $exp$ 0.5 & 4096 & 20480 \\
pl87, pl88, pl89, pl90, pl91 & g1 & p & - & 0.5 & $rotating$ & 4096 & 20480 \\
pl92, pl93, pl94, pl95, pl96 & g4 & p & - & 0.875 & $cold$ 0.25 & 4096 & 20480 \\
pl97, pl98, pl99, pl100, pl101 & g0 & p & p & 0.875 & $cold$ 0.25 & 1635 & 65400 \\
pl107, pl108, pl109, pl110, pl111 & g4 & p & - & 0.875 & $cold$ 0.5 & 4096 & 20480 \\
pl112, pl113, pl114, pl115, pl116 & g0 & p & p & 0.875 & $cold$ 0.5 & 1635 & 65400 \\
pl122, pl123, pl124, pl125, pl126 & g0 & p & p & 0.875 & $exp$ 1. & 1635 & 65400 \\
pl127, pl128, pl129, pl130, pl131 & g4 & p & - & 0.875 & $exp$ 1. & 4096 & 20480 \\
\hline
\hline
\end{tabular}
\end{center}
\end{minipage}
\end{table*}

The principal characteristics of the simulations discussed in this paper are 
listed in Table 3. Many more simulations were run than those listed. Since, 
however, they did not add new substantial information, or only corroborated 
results already obtained with other simulations, they will not be listed or 
discussed here. 
The first column gives the names of the runs, the second the
model used for the galaxy and the third the model used for the group. The
fourth gives the model for the common halo, a dash indicating that
there was no common halo, and the fifth column gives the ratio of halo-to-total
mass. The sixth one gives information on the kinematics of the initial 
configuration. $Virial$ denotes groups for which $2T/|W|=1$ and $cold$ groups 
with $2T/|W|=0.25$ or 0.5, in both cases the direction of the velocity vector 
being 
chosen at random. $Rotating$ groups also have $2T/|W|=1$, but now the 
component of the velocity perpendicular to the $z$ axis has been put in a 
direction perpendicular to the distance of the galaxy from the $z$ axis. 
Finally 
$exp$ denotes groups expanding with a uniform Hubble flow. The seventh column 
gives the number of particles in 
each galaxy and the eighth one
the total number of particles in the simulation. The number of particles in
each component has been chosen such that the mass of the particles constituting
the luminous part is equal to that of the particles constituting the halo. 
Missing
identification numbers refer to runs which are not essential for the 
discussions in this paper.

\subsection{Numerical miscellanea}
\label{sec:nummisc}
\indent

In our units $G=1$ and the total mass of the group is equal to 20 for all
models. In the case of groups with no common halo
the binding energy of the group, excluding the internal binding energy of the
galaxies, is -5. This is also approximately true for the cases with common 
halo.
Therefore, the velocity dispersion within a group is
$\sqrt{2|E|/M} = 1/\sqrt{2}$, where $E$ and $M$ are the binding energy and the
total mass of the group. 

Most of the numerical simulations have been carried out using direct 
summation on the 
GRAPE-3A and GRAPE-3AF 
boards in Marseille Observatory and a few on similar boards in Tokyo
University (for description of GRAPE 3 boards see e.g. Okumura \tal 1992, 
Ebisuzaki
\tal 1993 and Okumura \tal 1993). In the Marseille 
3AF configuration one time-step
takes 6.3 secs for simulations with
65400 particles. The time-step chosen is equal to 0.015625, and the softening 
in most cases equal to 0.03125. A few cases (namely runs King7 to King11 and
King27 to King29) have been evolved with a softening of 0.025, but that makes
no difference to the evolution. These parameters ensure an adequate energy
conservation. Thus on average the energy was conserved better than 0.1\% up 
to $t=100.$, better than 0.15\% up to $t=500.$ and better than 0.35\% up to 
$t=1000.$ (i.e. 64~000 time-steps). 

A number of the simulated groups took a long time before merging, and a few 
did not merge at all. It is thus of interest to ask whether the 
number of particles we use is sufficient to prevent our 
results from being severely influenced by two-body relaxation effects. The 
run which is most liable to be problematic is ki33 since it is modelled with 
16350 particles, the lowest we have used in our simulations, and furthermore 
has not merged by $t=1000.$ We thus reran this simulation with 
double the number of points (ki57) and compared the positions of galaxies as a 
function of time. Despite the length of the runs the positions of the galaxies 
in the two simulations agree on average better than 5\%.

\section{Calculation of merging rates}
\label{sec:rates}

\indent
To calculate merging rates 
we must first adopt a criterion determining when two galaxies merge. 
Barnes (1985) and
Bode, Cohn \& Lugger (1992) use a friends-of-friends algorithm, but this 
decides
that two galaxies have merged from the moment their halos or their luminous 
parts (depending on whether we apply it to the total or only to the
luminous mass distribution) have started to interpenetrate. 
Since parts of galaxies may overlap without them having merged, we believe 
that this criterion is not stringent enough, and we chose to 
associate
merging with loss of identity. 
Thus two galaxies are considered merged
if the distance between their centers is smaller than a given fraction of some 
characte\-ristic
radius, and the difference between their bulk velocities
smaller than a given fraction of some associated characteristic velocity.

\begin{figure*}
\includegraphics{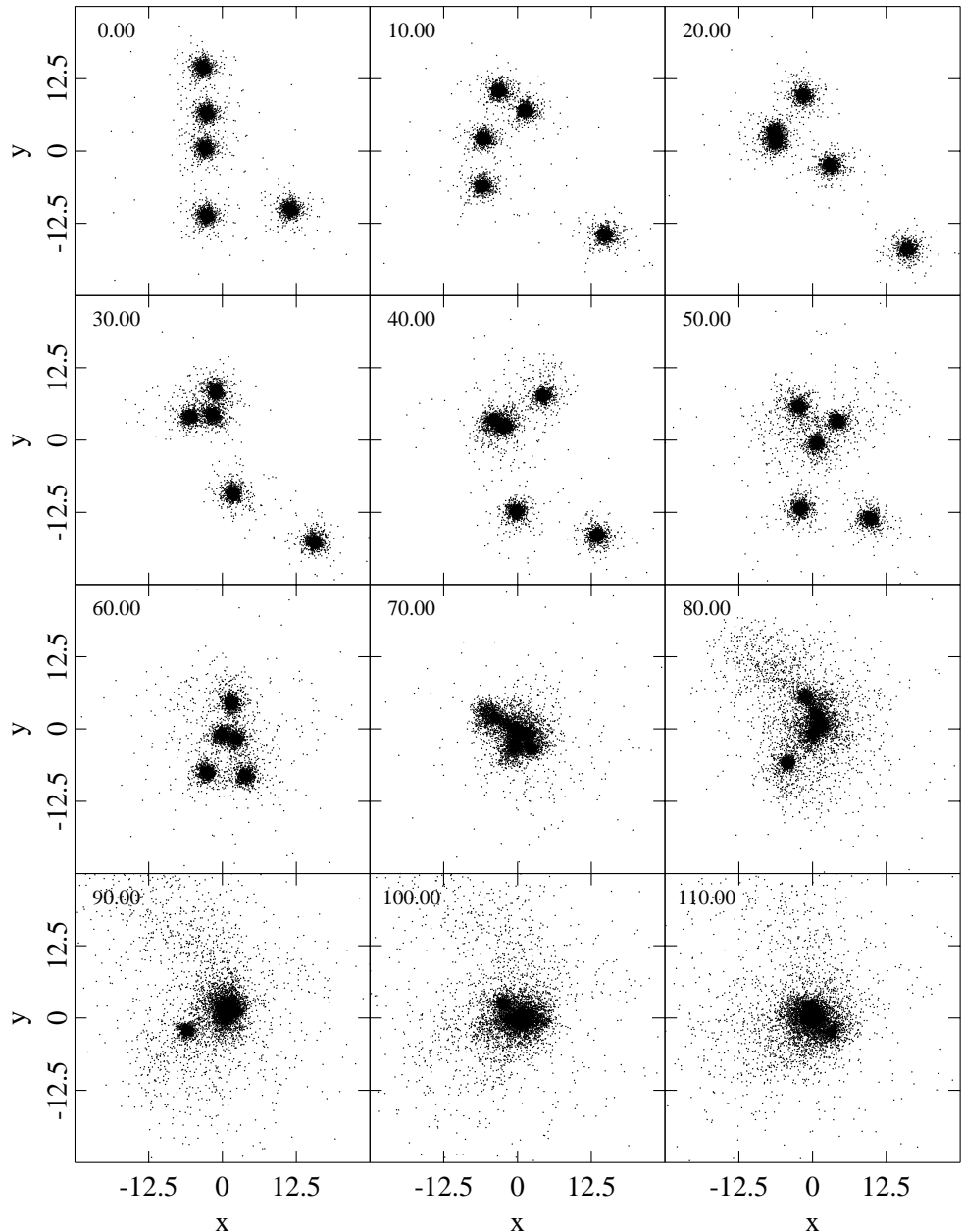}
\vspace{17.5cm}
\caption{Evolution of model ki16. Time increases from left to
right
and from top to bottom and is given in the upper left corner of each
frame. We plot the $xy$ projection of all particles representing the luminous 
matter.
}
\label{evol}
\end{figure*}

\begin{figure}
\includegraphics{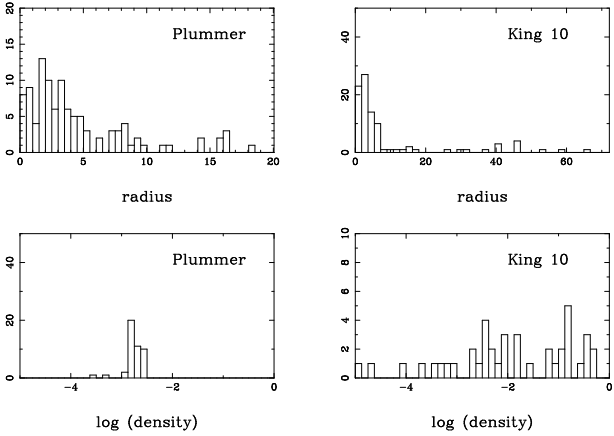}
\vspace{6.8cm}
\caption{Information on where the first, second and third merging occurred. 
The upper panels give histograms of the distances of these locations from the 
center-of-mass of the group. The two lower ones give histograms of the 
density of the unperturbed common halo at the merging locations. The left 
panels correspond to Plummer groups and the right panels to King $\Psi=10$ 
groups.}
\label{histo}
\end{figure}

We use
as comparison radii and velocities the minimum of the radii and
velocity dispersions of the two galaxies, so that the adopted criterion reads: 

\begin{equation}
R_{ij} < 0.5 min(r_{g,i}, r_{g,j})
\label{eq:crit1}
\end{equation}
\begin{equation}
V_{ij} < 0.5 min(\sigma_i, \sigma_j)
\label{eq:crit2}
\end{equation}
where $R_{ij}$ and $V_{ij}$ are the relative distance and velocity between
galaxies $i$ and $j$,  and $r_{g,k}$ and $\sigma_k$ are the  radius and central
velocity dispersion of galaxy $k$. The center of each galaxy is defined as the
position of the highest density, i.e. the average position weighted by the
local density, or 
\begin{equation}
{\bf r}_c = {\sum_{i=1}^N\rho_i {\bf r}_i \over
	    \sum_{i=1}^N\rho_i}
\label{eq:galcen}
\end{equation}
where ${\bf r}_i$ is the position of particle $i$, $\rho_i$ is the local
density around particle $i$ calculated using the six 
nearest neighbours and the sums are over all
particles that initially belong to the galaxy.
Similarly the characteristic radius of a galaxy is defined by: 
\begin{equation}
r_g = \sqrt{\sum_{i=1}^N\rho_i r_i^2 \over
	    \sum_{i=1}^N\rho_i}
\end{equation}
where $\rho_i$ and the summation are defined as above, and $r_i$ is the
distance of particle $i$ from the density center of each galaxy defined in eq. 
(\ref{eq:galcen}). Our
definition is similar, but not identical, to that used by von Hoerner (1963)
and Casertano and Hut (1985), since they use the average of the absolute 
distance
weighted by the local density to calculate the radius, while we calculate the
weighted root-mean-squared distance from the center. For the central velocity
dispersion we take the root-mean-squared velocity of particles within a radius
equal to twice $r_g$. 

Roughly speaking, our procedure to find the size and the velocity dispersion of
a galaxy is to look at the distribution of particles that are originally in
that galaxy, mark the region with highest density and calculate the
velocity dispersion within that region. To determine whether two galaxies have
merged, we check if these high-density regions overlap sufficiently and
if their relative velocity is small enough.  Note that our procedure can be
easily applied to galaxies which have already merged with other galaxies, since
it does not use the binding energy. We were thus able to use
this procedure to determine if all the five galaxies have formed one merger or
not, just by looking at all the pairs. 

\begin{figure}
\includegraphics{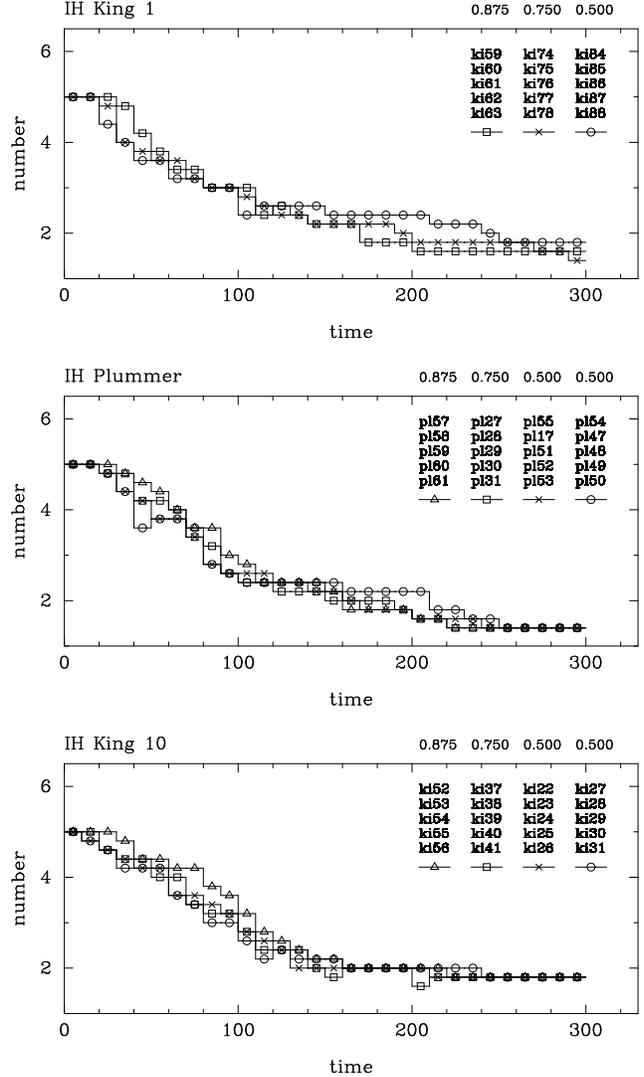}
\vspace{15.2cm}
\caption{Number of galaxies as a function of time for groups with individual
halos. The upper panel corresponds to King $\Psi=1$ groups and shows the mean
number of galaxies for groups with g1 galaxies (circles), groups with g3 
galaxies (crosses) and groups
with g4 galaxies (squares). The middle panel corresponds to Plummer groups
and shows the mean number of galaxies for groups with g1 galaxies (circles), 
groups with g2
galaxies (crosses), groups with g3 galaxies (squares) and groups
with g4 galaxies (triangles). The lower one corresponds to
King $\Psi=10$ groups and shows the mean number of galaxies for groups with g1
galaxies (circles), groups with g2 galaxies (crosses), groups
with g3 galaxies (squares) and groups with g4 galaxies (triangles). 
The merging rates of the different groups can hardly be distinguished from 
each other, and this
independent of the model of the group (King or Plummer). 
}
\label{mr_ih}
\end{figure}

This criterion performed very well in most cases. In only about 2\% 
of the cases it came to unreasonable conclusions, saying e.g. that galaxy 1 
has merged with galaxy 2 and galaxy 3, but galaxies 1 and 3 had not merged 
between them. In yet fewer 
cases, less than 1\%, the criterion declares two galaxies as having merged at 
a given time step, and not having merged in the next one. Such cases can be 
identified as 
temporary increases of the number of galaxies in Figures \ref{mr_ih} to 
\ref{mr_concen} and \ref{mr_rot} to \ref{fig_cold}.
This criterion was applied to every one of our runs and the results are
compared in figures \ref{mr_ih} to \ref{mr_concen} and \ref{mr_rot} to 
\ref{fig_cold}, which give the mean number of 
galaxies left in the group as a function of time. Since in all cases we 
average over 5 configurations, the
number of galaxies is of course not be necessarily an integer.
We repeated the exercise by applying the merging criterion to only the 
luminous part of the galaxies : our results for the merging rates
stay the same independent of whether we
consider the luminous part or the whole galaxy.

\section{Results}
\label{sec:results}

The evolution of a typical run is shown in Figure \ref{evol}. 
The galaxies move within the group and interact. Part of the kinetic energy 
of their bulk 
motions is converted to internal motions and they suffer orbital decay. When
they approach each other sufficiently close and with not 
too high relative velocity, they merge, until the group is reduced to only 
one object. 
The speed at which these mergings occur is to a large extent dependent on 
chance, via the randomly generated initial positions and velocities of 
the galaxies, but is 
also dependent on the global characteristics of the set-up, like the amount 
and distribution of dark matter, and in the following 
we will attempt to find out which factors determine this rate. In order to 
compensate to some extent for the element of chance every configuration was 
realised five times.

The positions where the mergings occur depend a lot on the initial
configuration in the sense that they reflect the initial distribution of
galaxies within the group, as is shown in Figure \ref{histo}. In order to 
obtain this
figure we measured for each simulation the distance of the locations where 
the first, second and
third mergings occurred from the center of mass of the group. The upper left
panel of Figure \ref{histo} shows the number of mergings as a function of 
that distance
for all simulations with a Plummer distribution starting off in virial 
equilibrium in (whenever relevant) a
Plummer common halo. The upper right panel has the same information, but now 
for King $\Psi=10$ distributions starting off in virial equilibrium in 
(whenever relevant) King $\Psi=10$ common halos. We note 
that in King $\Psi=10$ distributions a lot of the mergings occur either very
near the center or very far out, while in a Plummer distribution there are
many more mergings at intermediate distances. This difference is also 
reflected in the lower panels of the same figure, where instead of the 
distance from the center of mass we have the density of the unperturbed 
common halo at the location of the merging. Thus these panels do not contain 
simulations with individual halos. In the Plummer case the densities are all 
concentrated in a narrow region of values, while in the King $\Psi=10$ case 
both very 
high and very low densities are not uncommon. It seems that the distribution 
of galaxies in the group is the dominant factor in this case, and not the 
distribution of dark matter in the common halo, since mergings at very large 
distances from the group center of mass are also seen in the evolution of 
King $\Psi=10$ groups in Plummer halos (kp series of runs), although our 
statististics are much poorer in this case.

Somewhat less than 90\% of the mergings involve 2 galaxies, roughly 10\% 
involve 3 galaxies and only about 1\% involves 4 galaxies. In one case all 
5 galaxies of the group merged in the same time-step. 90\% of the pairs have 
a mass ratio 1:4 and 10\% have a mass ratio 2:3. Roughly 75\% of all triplets 
have a mass ratio 1:1:3, the remaining having a mass ratio 1:2:2.

\section{Comparing merging rates}
\label{sec:crates}

\subsection{Individual halos }
\label{sec:ihmass}
\indent

	To compare merging rates for groups with individual
halos of different masses and extents we plot in Figure \ref{mr_ih}  
the number of galaxies left at any given time as a
function of time. Surprisingly, the rates depend very
little on the model galaxy used, even though 
the relative mass and the radial extent of
the halos for galaxies g1, g2, g3 and g4 are quite different. This result
holds for all types of groups tested, i.e. Plummer groups and King $\Psi=$1, 
5 or 10 groups. Table 1 shows that both the radii containing 30\% and those 
containing
80\% of the total galaxy mass are about a factor 3.5 larger for model g4 than 
for model g1. 
On the other hand, since the mass in each configuration, and therefore each 
galaxy, is the same, more extended configurations have a smaller absolute 
value of the binding energy and therefore a 
smaller value of $\sigma_{gal}/\sigma_{group}$ (cf columns 4 and 5 of 
Table 1). 
Therefore galaxy encounters 
happen at a smaller (relative to their internal dispersion) velocity in cases 
with less extended halos, and this will induce higher merging rates. 
Thus two counterproducing effects are at work : when galaxies become more 
extended their merging rates increase, but, since at the same time their 
internal velocity dispersion decreases, this decreases their merging 
rates as well. 
It is not possible to decouple the two effects, unless we consider a 
different number of galaxies at the start, or a different total mass, or 
galaxies that start out of equilibrium, all three alternatives being 
undesirable. 
For our simulations 
the two effects must be of about equal importance since the merging 
rates do not depend in a systematic way on the model used. This can be seen 
in a more quantitative, albeit still very simple, way, as follows:

The probability that a given galaxy merges with another one in 
a unit time interval is expressed as
\begin{equation}
P = 4\pi \int_0^{\infty} f(v) n \sigma(v)v^3dv,
\end{equation}
where $f(v)$ is the distribution function of the relative velocity, $n$
is the number density of galaxies and $\sigma(v)$ is the merging cross section 
for a given relative velocity. The cross section $\sigma$ is
actually also a function of the type of the galaxies. For
simplicity, we here assume that galaxies are homologous, and therefore that
the relation
\begin{equation}
r\sigma_{gal}^2 = {\rm const.}
\end{equation}
holds for different galaxies, where $r$ is the characteristic radius
of the galaxy (for example the virial radius) and $\sigma_{gal}$ is its
velocity dispersion. If we write the cross section explicitly as a
function of $v$ and $r$, we have, from dimensional analysis:
\begin{equation}
\sigma(v,r) = k^{-2}\sigma(vk^{-1/2},kr).
\label{eqn:sigma-v-relation}
\end{equation}
Let us consider two groups consisting of galaxies
of two different sizes, $r_1$ and $r_2=k r_1$ respectively. We have
\begin{eqnarray}
P_1 &=& 4\pi\int_0^{\infty} f(v)\sigma(v,r_1)v^3dv,\nonumber\\
P_2 &=& 4\pi\int_0^{\infty} f(v)\sigma(v,r_2)v^3dv\nonumber\\
 &=& 4\pi k^{2}\int_0^{\infty} f(v)\sigma(vk^{1/2},r_1)v^3dv\\
 &=& 4\pi \int_0^{\infty} f(vk^{-1/2})\sigma(v,r_1)v^3dv,
\label{eqn:probability-eq}
\end{eqnarray}
There are two limiting cases. In one case, the velocity dispersion
within the cluster is much larger than the internal velocity
dispersion of galaxies. Since the cross section becomes zero at $v\sim
\sigma_{gal}$ (Makino and Hut 1996), we can, in this case, consider $f(v)$ to 
be independent of
$v$, unless the velocity distribution function is singular at
$v=0$. We then have
\begin{equation}
P_2 \sim P_1 .\quad \quad  (\sigma_{group}>> \sigma_{gal})
\end{equation}
In other words, in the limit of  $\sigma_{group}>> \sigma_{gal}$, the
merger rate, and therefore the lifetime of a group, does not depend on 
the size of the individual galaxies.

The other extreme is when the internal velocity dispersion of the galaxies is 
much
larger than that of the cluster. In this case we need an asymptotic
form of the cross section $\sigma$. Both numerical scattering
experiments and approximate theory suggest that $\sigma \propto
v^{-8/3}$ for small $v$ (Makino \& Hut, 1996). The gravitational focusing 
effect alone requires the power index to be larger than $-2$ (Makino \& Hut 
1996).  Thus, we have
\begin{equation}
P_2 \sim   k^{2/3}P_1 \quad \quad  (\sigma_{group}<< \sigma_{gal}),
\end{equation}
i.e. a weak dependence on $k$. Our simulations are
of course intermediate between these two extreme cases. 
Furthermore the above analysis is
valid only for the merging of two ga\-la\-xies in a cluster 
consisting of a large number of galaxies. Ne\-vertheless, the
above theoretical analysis, albeit very simplified, gives some insight to our 
result
and argues that our result is not surprising and that
the lifetime should depend only weakly  on the size of the galaxies.

\subsection{Common halos }
\label{sec:chmass}
\indent

\begin{figure}
\includegraphics{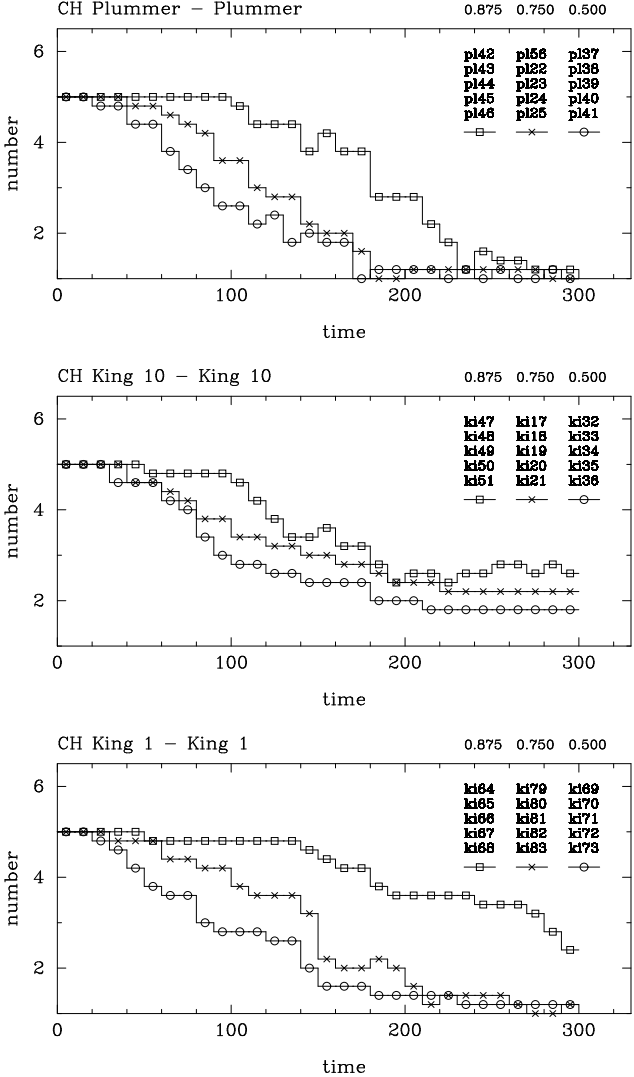}
\vspace{15.2cm}
\caption{Number of galaxies as a function of time for groups with common
halos. The upper panel corresponds to Plummer groups and common 
Plummer halos, 
and shows the mean
number of galaxies for groups with 0.5 (circles), 0.75 (crosses) 
and 0.875 (squares) halo-to-total mass ratio. The middle panel 
corresponds to King $\Psi=10$ groups in 
common King $\Psi=10$ halos and the lower one to King $\Psi=1$ groups 
in King $\Psi=1$ common halos. The 
different symbols correspond to the same halo-to-total mass ratio. 
}
\label{mr_ch1}
\end{figure}

\begin{figure}
\includegraphics{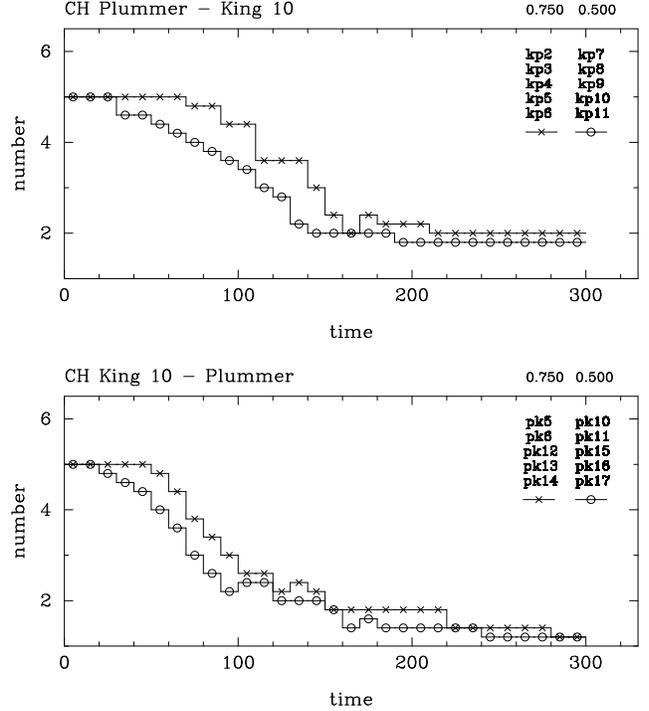}
\vspace{10.5cm}
\caption{As for the previous figure, but for King $\Psi=10$ groups 
in Plummer common halos (upper panel) and for Plummer groups 
in King $\Psi=10$ common halos (lower panel). The 
different symbols correspond to the same halo-to-total mass ratio (circles
for 0.5 and crosses for 0.75). It
is clear from this and the previous figure that groups with relatively more 
massive common halos merge slower 
than groups with less massive common halos, and that
independent of the model of the group or halo (King or Plummer). 
}
\label{mr_ch2}
\end{figure}

	Figures \ref{mr_ch1} and \ref{mr_ch2} compare merging rates for groups with common halos and different percentages of halo masses. We note that the 
higher the halo mass, the slower the merging will occur and this independent 
of whether the distribution of the galaxies and the halo are Plummer,   
King $\Psi=10$, or King $\Psi=1$. It is also true for mixed distributions, 
i.e. King $\Psi=10$ groups in Plummer 
halos, or Plummer groups in King $\Psi=10$ halos. This result can be best 
understood if we think of the galaxies as a perturbation on the global halo 
potential. If the halo-to-luminous mass ratio is small, the perturbations will 
be larger, the galaxies will be strongly attracted by each other and will 
merge faster. The opposite will be true for the case of large halo-to-luminous 
mass ratios. In the limiting case where the galaxies are test particles in 
the common halo they will merge only if their mutual trajectories 
intersect accidentally.

We can reach the same conclusion using arguments similar to those of the 
previous section. Let us again consider two groups consisting of galaxies 
with the same
size and with different masses $m_1$ and $m_2=km_1$ respectively. We then have
\begin{equation}
\sigma(v,m_2) = \sigma(v k^{-1/2},m_1),
\label{eqn:sigma-m-relation}
\end{equation}
instead of (\ref{eqn:sigma-v-relation}). In this case, following a
derivation similar to that in the previous section, we have
\begin{equation}
P_2 \sim \cases{ P_1 k^2, & $(\sigma_{group}>> \sigma_{gal})$\cr
 P_1 k^{4/3}, & $(\sigma_{group}<< \sigma_{gal})$}
\end{equation}
Thus, in both limits, the lifetime should depend relatively strongly on the 
mass of the
galaxies and in the same sense as found by the numerical simulations. 

\subsection{Individual versus common halos}
\label{sec:icmass}
\indent

\begin{figure}
\includegraphics{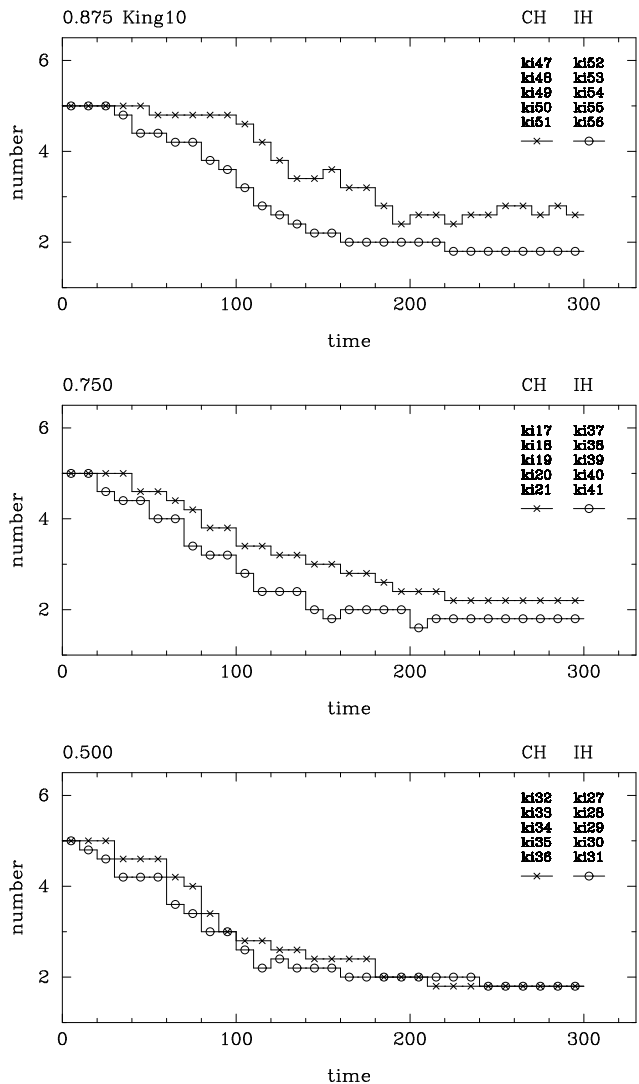}
\vspace{15.2cm}
\caption{
Number of galaxies as a function of time for King $\Psi=10$ groups. The upper 
panel 
corresponds to a halo-to-total mass ratio of 0.875, either in a common halo 
(crosses), or in individual halos (squares). The middle panel 
corresponds to a halo-to-total mass ratio of 0.75 and the lower one is for 
a halo-to-total mass ratio of 0.5, while the  
symbols are the same as for the upper panel. We note that the merging rate 
for models with individual halos is always higher than that for models with 
common halos, the difference being larger with increasing halo-to-total mass 
ratio.
}
\label{mr_ch_ih_k10}
\end{figure}

Fig. \ref{mr_ch_ih_k10} shows the behaviour of models with King distributions 
of index $\Psi=10$. 
The merging rate for
models with individual halos is higher than that for models with common
halos and this for all halo-to-total mass ratios and independent of the 
number of the remaining galaxies in
the group. 
This is in good agreement with the results of Barnes (1985) and Bode, Cohn \& 
Lugger (1992), who also use King distributions for the group and halo, albeit 
for $\Psi=7$, and different model galaxies.

\begin{figure}
\includegraphics{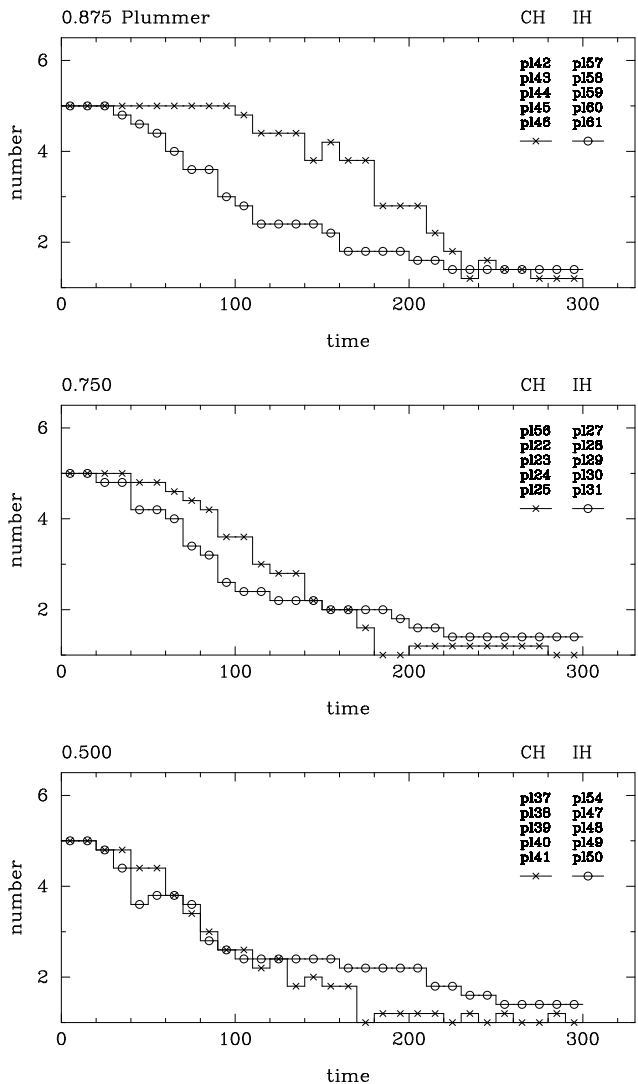}
\vspace{15.2cm}
\caption{
As for figure 6, but now for Plummer groups. We note that the evolution is 
faster initially for individual halo cases, then there is a cross-over and it 
is the common halo cases that merge faster. Where the cross-over occurs 
depends 
on the halo-to-total mass ratio.
}
\label{mr_ch_ih_pl}
\end{figure}

\begin{figure}
\includegraphics{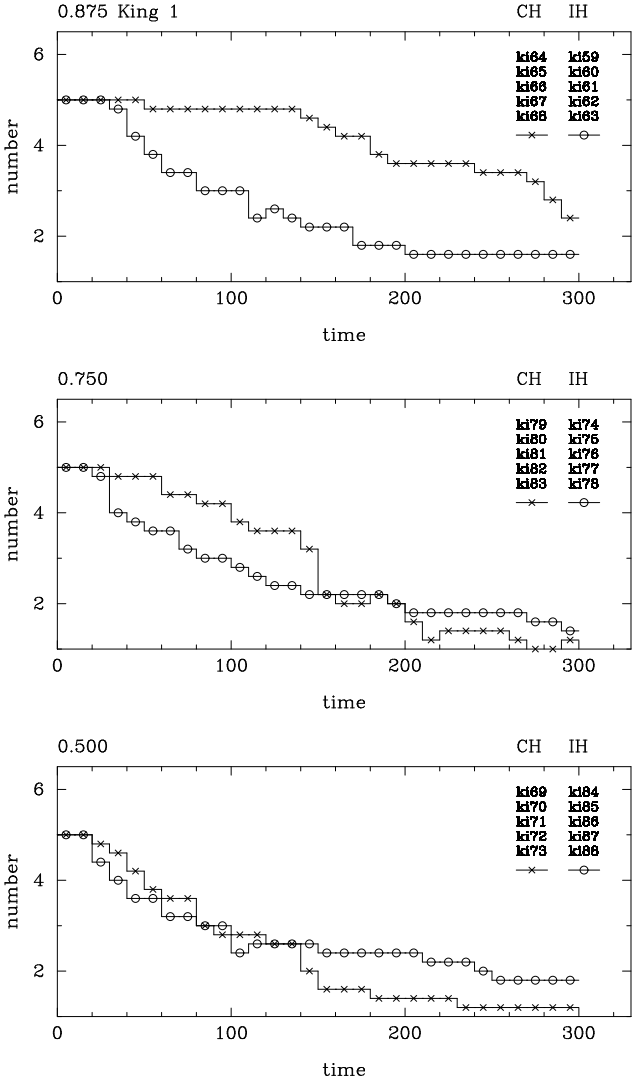}
\vspace{15.2cm}
\caption{
As for Figure 7, but now for King $\Psi=1$ groups.
}
\label{mr_ch_ih_k1}
\end{figure}

	Figure \ref{mr_ch_ih_pl} compares again merging rates of groups with 
common 
halos to groups with individual halos, but now for 
Plummer groups and, whenever relevant, Plummer common halos. 
In the upper panel, corresponding to a halo-to-total mass ratio of 0.875, we 
see that the merging goes consi\-de\-rably faster for groups with individual 
halos. The behaviour is more complicated for halo-to total mass ratio of 0.75 
or 0.5. Initially the number of mergers in the simulations with
individual halos is larger than that in the simulations with common
halos.  The trend, however, is reversed after some time
and configurations with common
halo show much larger merger rates than those with individual ones, due
to the fact that the configurations with common halos develop some
longer lived binary or triplet configurations than
the common halo cases. This cross-over occurs roughly when 2 ga\-la\-xies are 
left in the group for a halo-to-total mass ratio of 0.75 and at 3 galaxies 
for a 0.5 mass ratio. Since Athanassoula \& Makino (1995) were only discussing
the time of final total merge, these two panels show a behaviour in
good agreement with their results, which is not surprising since they
also used Plummer distributions.

Figure \ref{mr_ch_ih_k1} shows similar plots, but now for King $\Psi=1$ 
profiles. The
results are very similar to those found for Plummer distributions, in
the sense that for halo-to-total mass ratios of 0.75 and 0.5 the 
curves cross over, so that initially the merging rate is higher 
in the case of individual halos, the opposite being true for later
times. Furthermore the cross-over positions, in terms of galaxies 
left in the group, are roughly the same for Plummer and King $\Psi=1$ profiles.

There are two effects influencing the merging rate in an opposite sense. 
On the one hand in the case of individual halos the mutual attraction 
between galaxies is highest and that should favour faster merging.
On the other hand a dense common halo entails an important dynamical friction
and a corresponding slow-down of the galaxies, thus also favouring fast 
merging. It seems that the first effect dominates in the case of King $\Psi=10$
groups and in the initial stages of the evolution of Plummer and King $\Psi=1$
groups, and the second one in the remaining cases. This can be understood as 
follows: In the initial stages of the simulations the ga\-la\-xies are 
relatively 
far apart, therefore mutual attractions are important to help them come near 
each other and merge. Later in the evolution the half-mass 
radius of the luminous material has shrunk considerably and 
galaxies are very near each other. Thus mutual attractions are not
that important anymore, while dynamical friction slows the galaxies
down and speeds up the merging process. This, however, is not true for King
$\Psi=10$ models, since, as was discussed in section \ref{sec:results}, the 
distribution of the
merging positions reflects the initial distribution of the galaxies. 
Therefore for King $\Psi=10$ models several mergings occur relatively far 
from the 
center and the mutual attractions are still important to stimulate 
encounters.

\subsection{Effect of different distributions }
\label{sec:distri}
\indent

\begin{figure}
\includegraphics{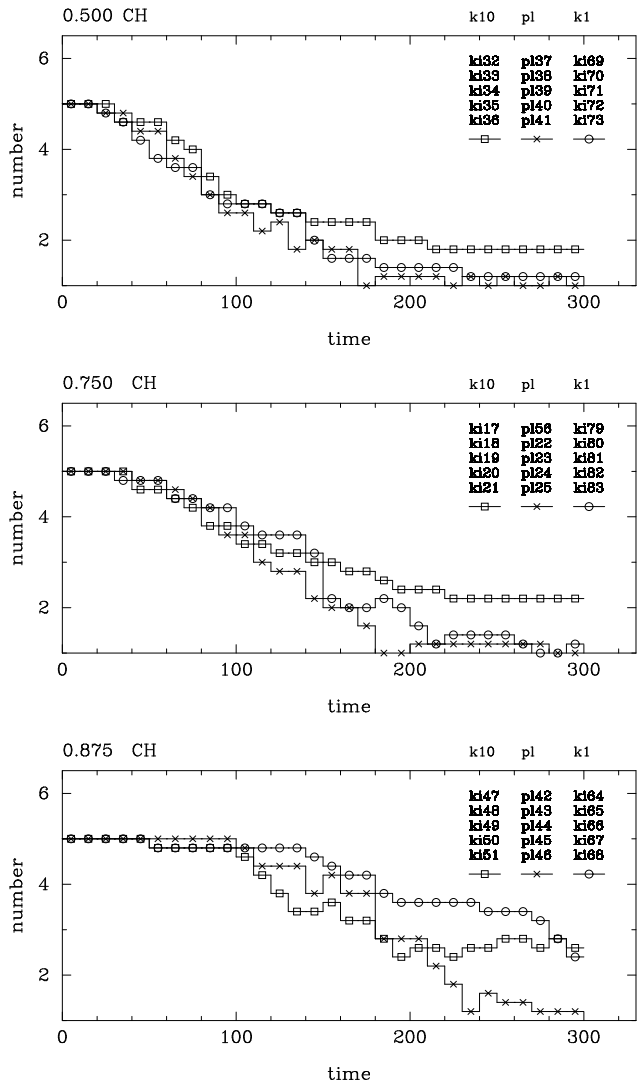}
\vspace{15.2cm}
\caption{
Number of galaxies as a function of time for simulations with common halos 
and a halo-to-total mass ratio of 0.5 (upper panel), 0.75 (middle panel) and 
0.875 (lower panel). The most centrally concentrated configuration (King 
$\Psi = 10$, squares), is compared to an intermediate one (Plummer, crosses) 
and to the least centrally concentrated (King $\Psi = 1$, circles). 
}
\label{mr_concen}
\end{figure}

	Let us now discuss the effect of central concentration on the merging 
rate. In the case of individual halos, comparing simulations with different 
distribution of the galaxies in the group, we see that there is a tendency 
for groups distributed in a more homogeneous way to merge faster during the 
last parts of the simulation. The effect, however, is very small. The 
comparison for cases with common halos can be seen in   fig. \ref{mr_concen}.
For low and intermediate halo-to-total mass ratios (upper and middle panels) 
the comparison goes as for si\-mu\-la\-tions with individual halos, although 
the effect of central concentration is somewhat stronger. The first few 
mergings happen quite early on and that independent of the central 
concentration. The last mergings occur much later in the case of King 10 
distributions than in the others. 

The effect of central concentration is most clearly seen  for the higher 
halo-to-total mass ratio (lower panel). It is the least centrally 
concentrated configuration, namely the King $\Psi = 1$ one, that merges the 
slowest. Since central concentration does not affect much the merging rates 
in simulations with individual halos, we infer that it is the central 
concentration of the common halo that makes the difference in this case. 
Then the observed behaviour can be explained as follows: 
 For the case of not centrally concentrated halos one can see that the 
galaxies oscillate in the common halo potential, influenced little by their 
mutual attractions, and slowed down by dynamical friction. One can consider 
such motions as damped oscillations, but with a relatively small damping 
factor, since the mass in the halo is distributed over a large volume so 
that the density, even for large halo-to-total mass ratios, is not very high. 
This is not the case for centrally concentrated halos. Now the dynamical 
friction is small when the galaxy is in the outer parts of the halo, and 
much bigger in the inner parts, particularly so when the galaxy has the high 
concentration parts behind it. In other words the galaxy is very heavily 
decelerated after it just crossed the center and, because of the energy it 
looses, it settles in the central parts of the potential well, so that 
mergings occur easier. Thus galaxies merge faster in more centrally 
concentrated common halos.

\begin{figure}
\includegraphics{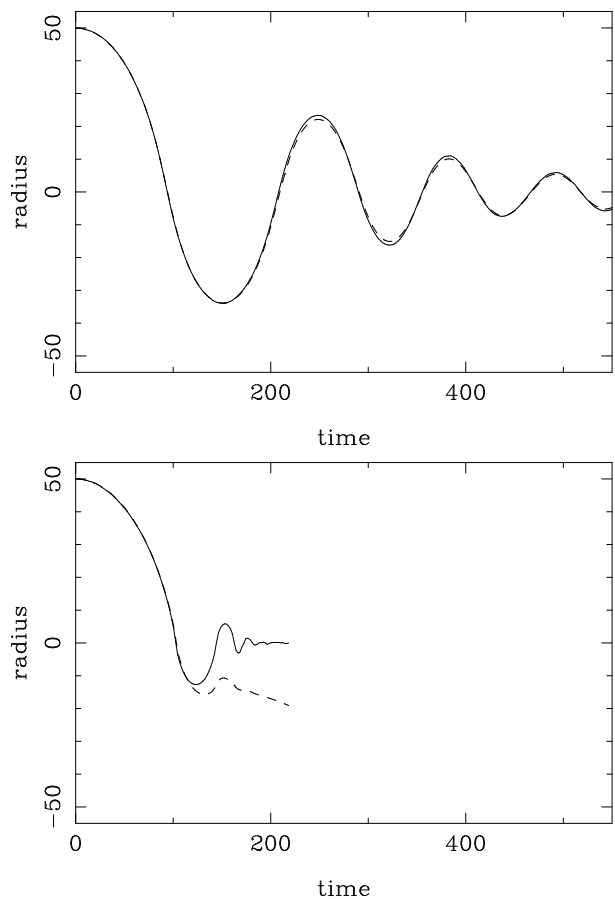}
\vspace{13cm}
\caption{
Distance of the center of a galaxy from the center of the common halo -
which is either a King $\Psi=1$ (upper panel), or a King $\Psi=10$ (lower 
panel). The center of the galaxy is obtained either from its center of mass 
(dashed line) or from its highest density point (solid line).
The simulation is described in the text.
}
\label{dyn_fr}
\end{figure}

To illustrate this we follow the motion of a galaxy in a common halo, which 
is either King $\Psi=1$ or  King $\Psi=10$. In both examples the galaxy 
starts off at $t=0$ from $x=50$ and has no initial velocity. The ratio of 
the galaxy-to-total mass is 0.025. In figure \ref{dyn_fr} we plot the d
istance of the galaxy from the center of the common halo as a function of 
time. The center of the galaxy can be either defined from its densest part 
(solid line), or its center of mass (dashed line). In the case of the King 
$\Psi=1$ common halo the galaxy does not suffer a serious deformation, nor 
does it loose much of its mass. Thus the maximum density point does not 
differ significantly from the center-of-mass. The galaxy oscillates around 
the center of the halo, while being slowly deccelerated by dynamical 
friction. It takes longer than $t=550$, or equivalently more than 3.5 
oscillations, before it is immobilised in the central part of the halo. 
The situation is very different in the case of a King $\Psi=10$ common halo. 
Now the galaxy suffers severe distortions when it goes through the halo 
center and looses a considerable fraction of its mass, starting from its 
first passage through the halo center. Thus the center of mass of the galaxy 
does not coincide with the point of maximum density. Following the latter, 
which is a more reasonable definition of the galaxy center, we see that the 
galaxy is decelerated in the central regions of the halo, looses an important 
fraction of its kinetic energy and can not reach the outer parts of the halo 
any more. This happens at relatively early times, before $t=200$. Staying in 
the inner parts it would, had other galaxies been presented in the 
simulation, be an easy prey to merging. Thus merging should proceed faster 
in centrally concentrated common halos.

\subsection{Different initial kinematics}
\label{sec:kinem}

\subsubsection{Cylindrically rotating groups}
\label{sec:rot}

\begin{figure}
\includegraphics{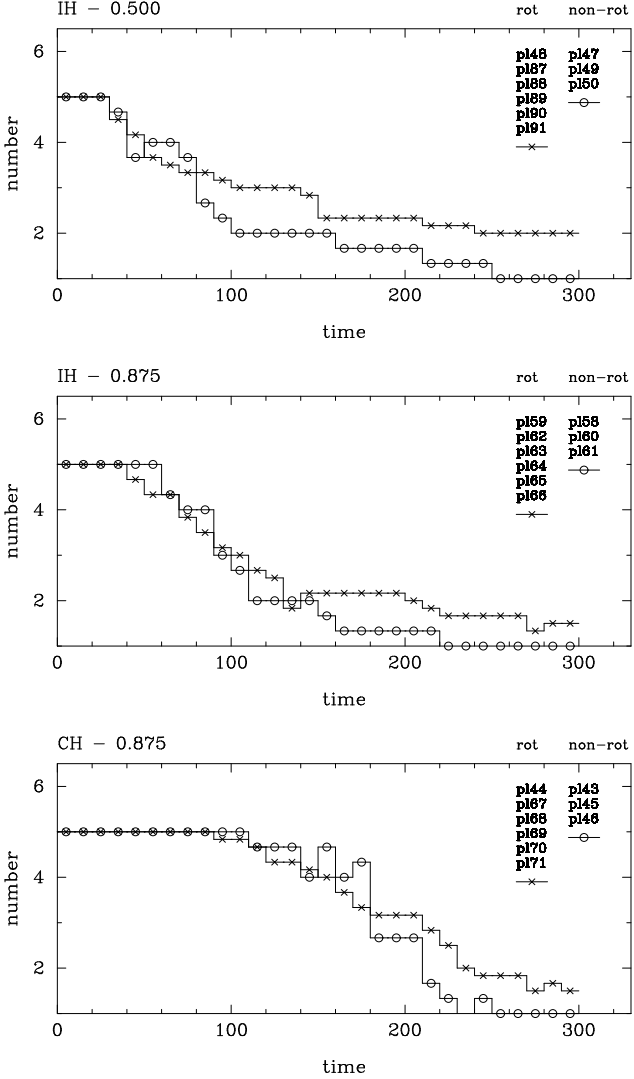}
\vspace{15.2cm}
\caption{
The effect of cylindrical rotation on the merging rate. The average of 
cylindrically rotating groups is given by crosses and of non-rotating ones 
by circles. The comparison is done for individual halo cases (upper two 
panels) and common halo ones (lower panel). The halo-to-total mass ratios 
considered are 0.5 (upper panel) and 0.875 (middle and lower panels).
}
\label{mr_rot}
\end{figure}

Our runs allow us to make three different comparisons between cylindrically
rotating and non-rotating groups, for different halo mass distributions and 
ratios, and we show them in Figure \ref{mr_rot}. The initial conditions for 
the cylindrically rotating groups were generated as described in section 
\ref{sec:simul}, namely by orienting the velocity vector of the mean velocity 
of each galaxy in the $(x,y)$ plane so that it is perpendicular to the 
cylindrical radius of the galaxy. These will be compared with initial 
conditions where this velocity vector is randomly oriented. It happened, 
however, that, out of the initial conditions generated with 
random orientation of the velocities, one had a substantial
amount of cylindrical rotation around the $z$-axis, so should be included
in the group of the fast $z$-rotators, while another one was an intermediate 
$z$-rotator, and thus was left out of the comparisons. We will thus be 
comparing in all
cases 6 groups of galaxies with fast cylindrical rotation around the
$z$-axis with 3 groups of galaxies with slow rotation. As shown in Figure
\ref{mr_rot} there is a clear indication that faster cylindrically rotating 
groups
take longer to merge than slower ones, as expected. The effect, however, is 
not very large.

\subsubsection{Cold groups}
\label{sec:cold}

\begin{figure}
\includegraphics{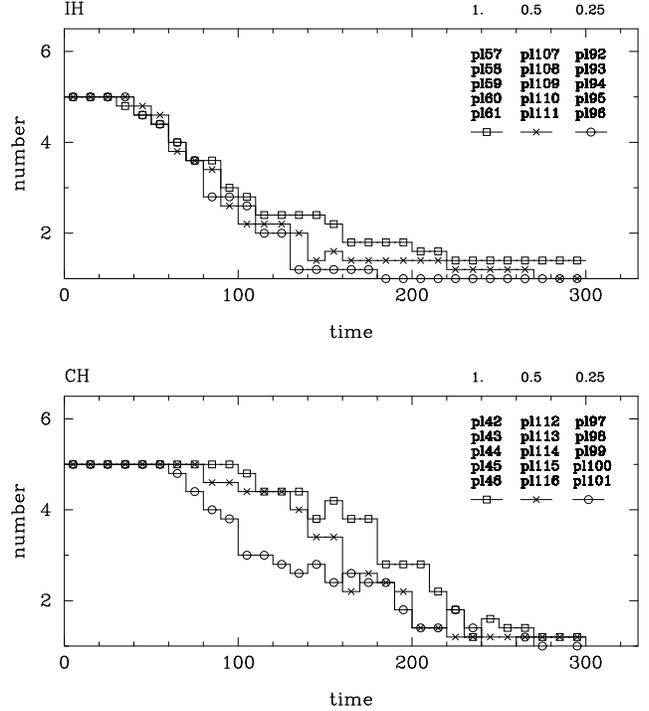}
\vspace{10.5cm}
\caption{
Number of galaxies as a function of time for simulations starting with 
$2T/|W|=0.25$ (circles), with $2T/|W|=0.5$ (crosses) and with $2T/|W|=1$ 
(squares). 
The upper panel corresponds to simulations with individual 
halos and the lower one 
to simulations with common halos. The halo-to-total mass ratio is
0.875 in all cases.
}
\label{fig_cold}
\end{figure}

Cold groups are more extended and have a smaller velocity dispersion. They 
start out of equilibrium, so they initially undergo a rapid collapse and a 
heating. Eventually, after some oscillations, their virial ratio reaches 
values around unity. At that time, provided merging does not hinder 
comparisons, the bulk velocities of the individual galaxies seem to be 
somewhat smaller than for the galaxies initially in virialised groups.

Figure \ref{fig_cold} compares the number of galaxies left at a given time in 
groups in
virial equilibrium ($2T/|W|=1$) with that of cold groups ($2T/|W|=0.25$ and 
0.5). The upper panel corresponds
to groups with individual halos and the lower one to groups with common halos;
in all cases the halo-to-total mass ratio is 0.875. We see that cold groups 
merge 
faster than the ones in virial equilibrium, as could be expected since 
encounters happen at smaller relative velocities and the angular momentum of 
the galaxies is in general smaller (cf. preceding section). Comparing the two
panels we also note that the difference between virial and collapsing groups 
is more important in the case of common halos than in the case
of individual ones. 

In Fig. \ref{fig_cold} we plot the number of galaxies as a function of times, 
as we did in all other cases. If, however, we measured time not in computer 
units but in multiples of some suitably defined initial crossing time, the 
difference between cold and virialised groups would be more important. This 
comes from the fact that cold groups are more extended and have lower bulk 
galaxy velocities, i.e. have larger crossing times. 

\subsubsection{Expanding groups}
\label{sec:expand}

There is not much difference in the merging rates of expanding groups with 
individual halos from those of the corresponding non-expanding ones, except 
for the last stages of the evolution. In the last stages the final pair 
survives longer and can often be unbound. This is due to the fact that, 
due to the Hubble flow initial conditions, the galaxy that is furthest 
from the center has a substantial outwards velocity and takes some time 
before turning around, if it ever does. On the other hand the galaxies 
that are in the central areas have similar merging histories as in the 
non-expanding cases. The effect of a Hubble expansion would presumably 
have been more important if the galaxies were initially less concentrated 
and located more in the outer parts, but we have not run such cases.

Expanding groups with common halos merge faster than virialised groups
with common halos. This effect is probably due to our initial conditions. 
Indeed the initial configuration is not in equilibrium, and the common halo 
evolves very fast towards a more concentrated triaxial configuration. The 
galaxies respond to this - as well as to the instability of their own 
configuration - and find themselves focused towards the inner parts,  
particularly in the direction of the minor and median axes of the halo. 
Thus encounters and subsequent mergers are favoured and the merging rates 
are higher than in simulations starting off near equilibrium. Obtaining 
initial configurations which have a Hubble expansion and start near a stable 
equilibrium is beyond the scope of this paper.

\section{Pairs and triplets}
\label{sec:pairs}
\indent

	All the 35 simulations with galaxies distributed according to 
Plummer models listed in Table 3 have merged before $t=500$. 
On the other hand out of 45 King $\Psi=10$ configurations
8 become unbound triplets, while another 8 pairs and 1 quartet have not 
merged 
although we have
continued the simulations for very long times, roughly up to $t = 1000$. 
In all cases they 
come from two starting conditions, the second, which we will hereafter call
condition B, and the third which we will call condition C. The reason is 
that, as we saw in section \ref{sec:galmod},
the King $\Psi=10$ model has a rather extended outer halo, so that 
occasionally points
will be drawn relatively far from the others. This is the case for
configuration C, where the fifth galaxy is initially far from the other four.
For the four cases with individual halos this configuration ends up as an
unbound triplet, two of its members having roughly twice the mass of the 
third one. The si\-mu\-la\-tions starting with configuration C and having 
common halos neither merge by the end of the
simulations, nor do they become unbound, but end up as bound pairs, presumably
to merge at times much larger than 1000. 

	Configuration B is a similar case, where all five ga\-la\-xies are 
initially relatively far apart. For the four cases with individual 
halos the configuration ends 
up as an unbound triplet. Four of the simulations with common halos neither 
merge by the end of the 
simulations, nor do they become unbound, but three end up as bound pairs and 
one
as an bound quartet. Two of the pairs have a mass ratio of 3:2 and one of 4:1.
The fifth case with common halo merged fully at $t = 1030$.

Groups with King $\Psi=1$ or 5 distributions give no unbound final
cases, although 1 of the 15 King $\Psi=5$ cases and 3 of the 30 King $\Psi=1$ 
did not merge fully by $t=1000$.

\section{How to get compact groups which merge slowly}
\label{sec:discuss}
\indent

The aim of our simulations was, given a system of five galaxies with a given 
total mass and energy, to find which configurations have the lowest merging 
rates. We consider both cases with individual halos and cases where the halo 
is common, enveloping the whole group. In the latter cases the merging rate 
is slower for cases where the halo is more massive. On the other hand the 
mass of individual halos does not influence much the merging rates, due to 
the fact that all galaxies have the same mass, and so more extended ones have 
a smaller velocity dispersion. Groups with individual halos merge faster than 
groups with common halos if the configuration is centrally concentrated, like 
a King distribution of index $\Psi=10$. On the other hand for less 
concentrated configurations the merging is initially faster for individual 
halo cases, the reverse being true after part of the group has merged. In 
cases with common halo centrally concentrated configurations merge slower 
for low halo-to-total mass ratios and faster for high mass ratios. Groups 
which have initially a virial ratio which is less than one merge faster, 
and groups that have initially cylindrical rotation merge slower than groups 
starting in virial equilibrium.  

Led by the above, we have tried to find one simulation which would have very 
slow merging rates. For simplicity we have restricted ourselves to cases 
initially in virial equilibrium, though if we waive this restriction we can 
get even longer lived configurations. Our simulations led us to try a case 
with a high halo-to-total mass ratio, distributed in a common halo which is 
as homogeneous as possible. In order to build such a halo we followed the 
evolution of a truncated homogeneous sphere composed of 155325 particles 
for 30 time units, a time sufficient for quasi equilibrium to be reached. 
We added 5 galaxies - composed of 1635 particles each - with initial mean 
positions and velocities drawn at random from the particles constituting 
the common halo, rescaled  appropriately so that the simulation would start 
in equilibrium and represent a compact group in virial equilibrium and with 
a ratio of common halo mass to total mass of 0.95. We evolved this in the 
same way as all the other simulations. The striking result is that the first 
merging occurred only at $t=990$, followed closely by the subsequent mergings 
at $t=1010, 1060$ and 1110 respectively. This is considerably longer than 
corresponding times for other si\-mu\-la\-tions. (This simulation took 32 
days on the Marseille GRAPE-3AF system).
For comparison let us note that for simulations with individual halos 
starting off in virial equilibrium the mean times for the four mergings 
are roughly 60, 85, 145 and 243 respectively. If we use as length the 
radius containing 80\% of the halo mass at $t=0$ and as velocity the value 
$1/\sqrt{2}$, then we get a measure of the crossing time roughly equal to 34. 
This means that the four mergings in this simulation occurred after roughly 
29, 30, 31 and 33 crossing times. 

Converting times to astronomical units is neither straightforward nor unique. 
On the one hand observations give us estimates of the velocity dispersion i
n the group now, and not at the beginning of the evolution. Also, for a 
fairer comparison, the mass of the group should include the halo mass within 
the volume occupied by the galaxies at the beginning of the evolution and 
not now. Nevertheless we can try a rough estimate. Thus using for a group 
mass $10^{13} M_{\odot}$ and for the velocity dispersion 300 $km/sec$ we 
get that a 1000 computer units for time correspond roughly to 28 Gyrs. 
Considering a smaller mass for the group or a larger velocity dispersion 
would of course reduce this value. 

This value is sufficiently large to allow us to conclude that a compact 
group such as described in this section can survive for a time comparable 
or larger to the age of the universe. It should also be noted that the 
value of common-halo-to-total mass ratio used in this example is not 
excessive. Pildis, Bregman \& Evrard (1995) have analysed a sample of 12 
HCGs plus the NGC 2300 group. For those groups that have extended X-ray 
emission, i.e. approximately two-thirds of the sample, they find a baryon 
fraction between 5\% and 19\%. If we take into account that the hot gas 
component contributes part of the baryons, while in our analysis it would 
contribute to the common halo mass, we see that the mass fraction we 
adopted is not unreasonable. Furthermore, as we already mentioned, some 
more longevity could be obtained if we also use the most favourable 
kinematical initial conditions, rather than limiting ourselves to virial 
equilibrium by default.

Thus our simulations point to a third solution to the compact group problem, 
providing an alternative explanation for why we observe so many compact 
groups despite the fact that their lifetimes are believed to be so short. 
Namely it could be that the groups we observe have survived because they 
have a common halo, of considerable mass, distributed in an appropriate 
way, and/or have appropriate kinematical initial conditions. The other 
solutions proposed so far are that the compact groups are not real entities, 
but chance projections of at least some members of the group (e.g. Rose 
1979; Mamon 1986, 1987, 1995; Hernquist, Katz \& Weinberg 1994); and that 
they continuously form as subunits of rich groups (e.g. Diaferio, Geller 
\& Ramella 1994). In order to discuss seriously these three alternatives 
and their implications we first have to make a thorough analysis of the 
structure and kinematics of the merger remnants, which we will do in a 
future paper.
  
Finally let us note that our simulations have not been done within a 
cosmological
context and that, despite the large number of simulations tried (over 
250) many effects have not been addressed. Thus all our initial conditions 
have been drawn from distributions which are spherically symmetric, 
isotropic, isolated and devoid of gas, and all galaxies are equal mass 
spherical ellipticals. Some of these assumptions will certainly influence 
the merging rates and it could be that their effect could further prolong 
the lifetime of a group. Such a study, however, is beyond the scope of 
this paper.

\section*{Acknowledgements}

We would like to thank Jean-Charles Lambert for his help with the 
administration of the simulations. E. A. and A. B. gratefully acknowledge 
the hospitality of the College of
Arts and Sciences of the University of Tokyo and CNRS-JSPS exchange grants
which made their trips
possible. They would also like to thank the
INSU/CNRS and the University of Aix-Marseille I for funds to develop
the computing facilities used for the calculations in this paper.
J. M. gratefully acknowledges the hospitality of the Observatoire de 
Marseille and CNRS-JSPS exchange grants
which made his trips
possible. The NEMO package was often used both in the generation of the initial
conditions and the analysis of the simulations and we are indebted to Peter
Teuben for his efforts to maintain it.

\bsp

\label{lastpage}

\end{document}